\documentclass[doublecol,linenumbers]{epl2}
\usepackage{amsmath}

\title{Quantum Monte Carlo study of the $S_4$ symmetric microscopic model
for iron-based superconductors}

\author{Guang-Kun Liu\inst{1} \and Zhong-Bing Huang\inst{2,3} \and Yong-Jun Wang\inst{1}}
\shortauthor{G.-K. Liu \etal}
\institute{
  \inst{1} Department of Physics, Beijing Normal University - Beijing 100875, China\\
  \inst{2} Department of Physics, Hubei University - Wuhan 430062, China\\
  \inst{3} Beijing Computational Science Research Center - Beijing 100084, China
}


\abstract{The $S_4$ symmetric microscopic model with two iso-spin
components has been studied via constrained-path quantum Monte Carlo
simulation. Our results demonstrate a stable $(\pi,0)$ or $(0,\pi)$
magnetic order which is significantly enhanced on increasing both
the Coulomb repulsion $U$ and Hund's coupling strength $J$. Also,
our simulation indicates that the magnetic order tends to be in an
orthomagnetic one, in which the nearest-neighbour magnetic moment
are orthogonal to each other, rather than in a collinear
antiferromagnetic state. Interestingly, when the system is doped
away from half filling, the magnetic order is obviously elevated in
the low doping density, and then significantly suppressed when more
electrons are introduced. Meanwhile, we find that an $A_{1g}$
$s_{\pm}$-wave pairing dominates all the singlet nearest-neighbour
pairings, and is significantly enhanced via electron doping.}

\begin{document}
\maketitle

\section{Introduction}
Iron-based superconductors (IBSCs) have triggered lots of attentions
since they were discovered in 2008. Through years of intensive
studies, it is widely believed that the sign-reversing $s$-wave, so
called $s_{\pm}$-wave pairing state~\cite{mazin2008,chubukov2009},
is the most probable pairing symmetry for IBSCs. However, some
argues that $d$-wave~\cite{kuroki2008,graser2009} or
$p$-wave~\cite{lee2008,brydon2011} pairings are also possible
candidates. It seems to be a reasonable strategy to find out more
evidences of the exact pairing symmetry through theoretical models,
and indeed several initial multi-orbital
models~\cite{raghu2008,daghofer2010,kuroki2008}, constructed with 2
to 5 orbitals, have been proposed to understand IBSCs. However, most
researchers presuppose that models without considering all active
orbitals in IBSCs are insufficient~\cite{johnston2010}, which means
at least 5 orbitals should be included for a ``proper'' model.
Obviously, it is very hard for current theoretical approaches to
make reliable predictions.

Interestingly, with proper considerations of the $S_4$ symmetry in
FeX (X refers As or Se) trilayers, the building blocks of IBSCs, an
effective two-orbital model has been established and proven to
essentially capture the underlying low-energy physics of
IBSCs~\cite{hu2012}. Compared with other multi-orbital models for
IBSCs, the $S_4$ model not only builds possible connections between
the IBSCs and cuprates~\cite{hu2012,ma2013}, but also offers a
comprehensive and novel picture describing the complex kinematics in
IBSCs: Fe $3d_{xz/yz}$-orbitals are divided into two nearly
degenerate and weakly coupled groups (so called $S_4$ iso-spins),
which are properly linked with $S_4$ transformation. The kinematics
of each group and the hybridization between them constitute the
$S_4$ model.

Considering the weak coupling between the two components, it is
argued that the physics of only one $S_4$ iso-spin may capture the
main features of the model. So as a first order approximation, the
$S_4$ model can be further reduced to a single iso-spin one
described by an extended one-orbital Hubbard model near half
filling~\cite{hu2012,ma2013}. Because of its relative
simplification, most previous researches on $S_4$ model focus on the
single iso-spin case. Using a finite-temperature quantum Monte Carlo
(QMC) method, Ma~\etal~\cite{ma2013} have simulated the model on
square lattices and demonstrated a stable $(\pi,0)$ or $(0,\pi)$
antiferromagnetic correlation at half filling and a dominant
extended-$s$-wave pairing over other pairings at low temperatures;
while another ground-state QMC study has also confirmed this pairing
symmetry in various lattices and wide range of
parameters~\cite{wu2013}.

Few works concentrate on the full $S_4$ model with two iso-spins,
however, it would be of interest and importance to investigate how
the multi-orbital interactions, such as Hund's coupling and pairing
hopping, could influence the magnetic and pairing properties. In
this letter, using our recently improved constrained-path quantum
Monte Carlo (CPQMC) method for multi-orbital models~\cite{liu2014},
we systematically studied the magnetic order and the pairing
correlation of the two-orbital $S_4$ symmetric microscopic model. We
find a stable $(\pi,0)$ or $(0,\pi)$ magnetic order at half filling
for various Coulomb repulsion $U$ and Hund's strength $J$, which are
consistent with other multi-orbital models for
IBSCs~\cite{liu2014,moreo2009a,nicholson2012,daghofer2008}. The
magnetic order is obviously favoured at low electron doping and then
sharply suppressed when we keep on increasing the doping density,
which also agrees well with our previous QMC simulations of another
two-orbital model~\cite{liu2014}. Finally, we find that a
doping-assistant $s_\pm$-wave pairing symmetry dominates all the
pairing channels.

\begin{figure}
\onefigure[scale=0.4]{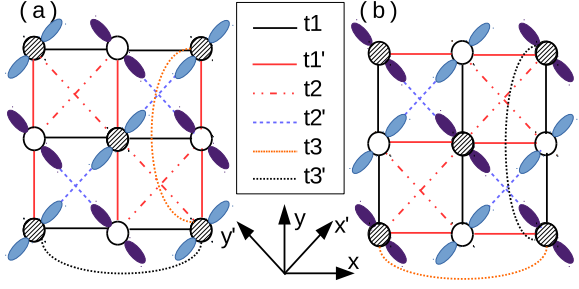} \caption{A sketch of the
$d_{x'z}$ and $d_{y'z}$ orbitals and schematic showing of hopping
parameters for each $S_4$ iso-spin components. Empty and hatched
circles represent different sublattice A and B, respectively. In
reality, the two iso-spins overlap completely as shown in fig.~3 of
ref.~\cite{hu2012}, we plot them separately for a better view of the
hopping parameters in each iso-spin. It is noted that the sign of
the hopping parameters are not reflected in the figure.}
\label{fig0}
\end{figure}

\section{Model and numerical approach}
Band calculations indicate strong hybridizations between Fe $3d$-
and As (Se) $p$-orbitals near the Fermi surface, and obviously
$d_{x'z}$ and $d_{y'z}$ have the largest overlaps with $p_x'$ and
$p_y'$ orbitals along the sublattice directions $x'$ and
$y'$~\cite{hu2012} (see fig.~\ref{fig0}). Meanwhile, considering
that the two As (Se) layers are separated apart along the $c$ axis,
the Fe $3d$-orbitals can be divided into two single-orbital
groups~\cite{hu2012,hao2013,hu2013}, as shown in fig.~\ref{fig0}:
One is consisted of $d_{x'z}$ on sublattice A and $d_{y'z}$ on
sublattice B, and these two obitals strongly couple to the
$p$-orbitals of the upper As (Se) layer. Comparatively, the other
group has $d_{y'z}$ on sublattice A and $d_{x'z}$ on sublattice B,
but couple to the lower As (Se) layer. These two iso-spins are
degenerate and weakly coupled, and can be mapped into each other via
$S_4$ transformation.

Based on these assumptions, the $S_4$ symmetric microscopic model
can be constructed as a combination of the kinematics of the two
iso-spins and the hybridization between them. Specifically, the
kinetic Hamiltonian of the $S_4$ model can be expressed
as~\cite{hu2012}

\begin{eqnarray}
H_{\rm kin}&=&H_{\rm kin}^1+H_{\rm kin}^2+H_{\rm kin}^c,\\
H_{\rm
kin}^1&=&t_1\sum_{\mathrm{i}\sigma}(a_{\mathrm{i},1,\sigma}^{\dagger}
                b_{\mathrm{i}+\hat{x},1,\sigma}+ \mathrm{h.c.}) \\
&&+t_1^{\prime}\sum_{\mathrm{i}\sigma}(a_{\mathrm{i},1,\sigma}^{\dagger}
                b_{\mathrm{i}+\hat{y},1,\sigma}+ \mathrm{h.c.})\notag \\
&&+t_2\sum_{\mathrm{i}\sigma}(a_{\mathrm{i},1,\sigma}^{\dagger}
                        a_{\mathrm{i}\pm(\hat{x}+\hat{y}),1,\sigma}
                       +b_{\mathrm{i},1,\sigma}^{\dagger}
                        b_{\mathrm{i}\pm(\hat{x}-\hat{y}),1,\sigma})\notag \\
&&+t_2^{\prime
}\sum_{\mathrm{i}\sigma}(a_{\mathrm{i},1,\sigma}^{\dagger}
                        a_{\mathrm{i}\pm(\hat{x}-\hat{y}),1,\sigma}
                       +b_{\mathrm{i},1,\sigma}^{\dagger}
                        b_{\mathrm{i}\pm(\hat{x}+\hat{y}),1,\sigma})\notag \\
&&+t_3\sum_{\mathrm{i}\sigma}(a_{\mathrm{i},1,\sigma}^{\dagger}
                        a_{\mathrm{i}\pm 2\hat{x},1,\sigma}
                       +b_{\mathrm{i},1,\sigma}^{\dagger}
                        b_{\mathrm{i}\pm 2\hat{x},1,\sigma})\notag \\
&&+t_3^\prime\sum_{\mathrm{i}\sigma}(a_{\mathrm{i},1,\sigma}^{\dagger}
                        a_{\mathrm{i}\pm 2\hat{y},1,\sigma}
                       +b_{\mathrm{i},1,\sigma}^{\dagger}
                        b_{\mathrm{i}\pm 2\hat{y},1,\sigma})\notag \\
H_{\rm
kin}^2&=&-t_1^\prime\sum_{\mathrm{i}\sigma}(a_{\mathrm{i},2,\sigma}^{\dagger}
                b_{\mathrm{i}+\hat{x},2,\sigma}+ \mathrm{h.c.}) \\
&&-t_1\sum_{\mathrm{i}\sigma}(a_{\mathrm{i},2,\sigma}^{\dagger}
                b_{\mathrm{i}+\hat{y},2,\sigma}+ \mathrm{h.c.}) \notag \\
&&-t_2^\prime\sum_{\mathrm{i}\sigma}(a_{\mathrm{i},2,\sigma}^{\dagger}
                        a_{\mathrm{i}\pm(\hat{x}+\hat{y}),2,\sigma}
                       +b_{\mathrm{i},2,\sigma}^{\dagger}
                        b_{\mathrm{i}\pm(\hat{x}-\hat{y}),2,\sigma})\notag \\
&&-t_2\sum_{\mathrm{i}\sigma}(a_{\mathrm{i},2,\sigma}^{\dagger}
                        a_{\mathrm{i}\pm(\hat{x}-\hat{y}),2,\sigma}
                       +b_{\mathrm{i},2,\sigma}^{\dagger}
                        b_{\mathrm{i}\pm(\hat{x}+\hat{y}),2,\sigma})\notag \\
&&+t_3^\prime\sum_{\mathrm{i}\sigma}(a_{\mathrm{i},2,\sigma}^{\dagger}
                        a_{\mathrm{i}\pm 2\hat{x},2,\sigma}
                       +b_{\mathrm{i},2,\sigma}^{\dagger}
                        b_{\mathrm{i}\pm 2\hat{x},2,\sigma})\notag \\
&&+t_3\sum_{\mathrm{i}\sigma}(a_{\mathrm{i},2,\sigma}^{\dagger}
                        a_{\mathrm{i}\pm 2\hat{y},2,\sigma}
                       +b_{\mathrm{i},2,\sigma}^{\dagger}
                        b_{\mathrm{i}\pm 2\hat{y},2,\sigma})\notag \\
H_{\rm
kin}^c&=&t_c\sum_{\mathrm{i}\eta\sigma}(a_{\mathrm{i},1,\sigma}^{\dagger}
 b_{\mathrm{i}+\eta,2,\sigma}+\mathrm{h.c.}),
\end{eqnarray}
where $a_{\mathrm{i},\alpha,\sigma}^{\dagger}$ ($a_{\mathrm{i},\alpha,\sigma}$)
creates (annihilates) an electron with spin-$\sigma$ at site $\mathrm{R_i}$ on
the sublattice A for the iso-spin $\alpha$ ($\alpha=1,2$), and similarly
$b_{\mathrm{i},\alpha,\sigma}^{\dagger}$ ($b_{\mathrm{i},\alpha,\sigma}$) acts on
sublattice B. The index $\eta=\hat{x}$ or $\hat{y}$ denotes a unit vector
linking the nearest-neighbour sites. Following ref.~\cite{hu2012}, the typical
hopping parameters for iron pnictides will always be chosen as $t_1=0.37$,
$t_1^\prime=0.43$, $t_2=0.90$, $t_2^\prime=-0.3$, $t_3=0.0$, $t_3^\prime=0.1$
and $t_c=0.02$ in our simulations.

The interaction Hamiltonian $H_{\rm int}$, containing a Hubbard
repulsion $U$ within the same iso-spin, a repulsion $U^{\prime}$ for
different iso-spins, a ferromagnetic Hund's coupling $J$ and
pair-hopping terms, can be written as

\begin{equation}
\label{eq2}
\begin{split}
H_{\rm int}=&J\sum_{\mathrm{i},\alpha\neq\alpha^\prime}
       (d_{\mathrm{i}\alpha\uparrow}^\dagger d_{\mathrm{i}\alpha^\prime\downarrow}^\dagger
        d_{\mathrm{i}\alpha\downarrow} d_{\mathrm{i}\alpha^\prime\uparrow} \\
       &+d_{\mathrm{i}\alpha\uparrow}^\dagger d_{\mathrm{i}\alpha\downarrow}^\dagger
        d_{\mathrm{i}\alpha^\prime\downarrow} d_{\mathrm{i}\alpha^\prime\uparrow}) \\
   &+(U^\prime-J)\sum_{\mathrm{i},\sigma}n_{\mathrm{i},1,\sigma}n_{\mathrm{i},2,\sigma} \\
   &+U\sum_{\mathrm{i},\alpha}n_{\mathrm{i}\alpha\uparrow}n_{\mathrm{i}\alpha\downarrow}
    +U^\prime\sum_{\mathrm{i},\sigma}n_{\mathrm{i},1,\sigma}n_{\mathrm{i},2,-\sigma},
\end{split}
\end{equation}
where $d_{i,\alpha,\sigma}^\dagger$ ($d_{i,\alpha,\sigma}$) creates
(annihilates) a spin-$\sigma$ electron at site $\mathrm{R_{i}}$
(sublattice A or B) for iso-spin $\alpha$ ($\alpha=1,2$), and
$U^\prime$ satisfies the constraint $U^{\prime}=U-2J$ due to the
rotational invariance~\cite{dagotto2001}.

\begin{figure}
\onefigure[scale=0.6]{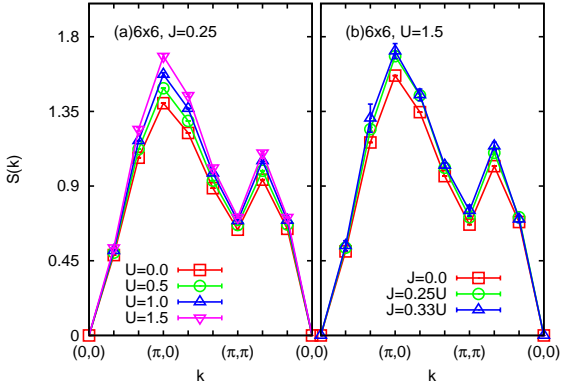}
\caption{Magnetic structure factor $S(k)$ at half filling on a 6$\times$6 lattice
versus various (a) $U$ and (b) Hund's coupling $J$.}
\label{fig1}
\end{figure}

\begin{figure}
\onefigure[scale=0.6]{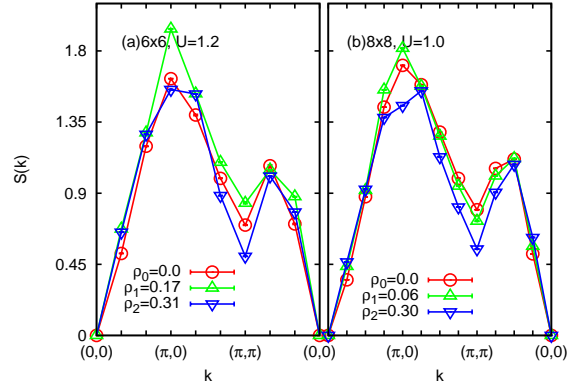}
\caption{(a) Magnetic structure factor $S(k)$ of  (a) $6\times 6$ and (b) $8\times8$
lattices on three typical electron doping densities.}
\label{fig2}
\end{figure}

We employ the CPQMC method~\cite{zhang1997a} to study the system. In
the CPQMC method, like other projector ground state QMC method, the
ground state, represented by a Slater determinant $|\phi_g\rangle$,
can be projected iteratively from any non-orthogonal, initial state
$|\phi_t\rangle$ via branching random walks in the overcomplete
Slater determinant space --- $|\phi^{n+1}\rangle= e^{-\Delta\tau
H}|\phi^{n}\rangle$ with $|\phi^0\rangle\equiv |\phi_t\rangle$  and
$H$ being the Hamiltonian. Differently, CPQMC requires every random
walker $|\phi^n\rangle$ in the iterations obey the restriction
$\langle\phi_t|\phi^n\rangle>0$. If the initial state happened to be
the ground state of the system, $|\phi_t\rangle=|\phi_g\rangle$, no
sign problem would ever appear under
$\langle\phi_t|\phi^n\rangle>0$~\cite{zhang1997a}. Obviously, such
an ideal situation never occurs in practical simulations. But even
under the approximate restriction $\langle\phi_t|\phi^n\rangle>0$,
CPQMC still efficiently eliminates the infamous Fermi sign problem
and obtains very high accurate results~\cite{zhang1997a,zhang1997b}.

In the usual CPQMC algorithm, before the projecting iteration
$|\phi^{n+1}\rangle= e^{-\Delta\tau H}|\phi^{n}\rangle$, we often
transform $e^{-\Delta\tau H}$ into combinations of simple items that
can be easily handled with, for example, we decouple the
$e^{-\Delta\tau Un_{i\uparrow}n_{i\downarrow}}$ into $e^{-\Delta\tau
U(n_{i\uparrow}+n_{i\downarrow})/2}\sum_{\sigma=\pm 1}
e^{\gamma\sigma(n_{i\uparrow}-n_{i\downarrow})}$ via discrete
Hubbard-Stranovich (HS) transformation~\cite{hirsch1983}. However,
considering the much more complex interaction items in the
two-orbital system, such as $H_1=J\sum_{\alpha\neq\alpha^\prime}
(d_{\mathrm{i}\alpha\uparrow}^\dagger
d_{\mathrm{i}\alpha^\prime\downarrow}^\dagger
 d_{\mathrm{i}\alpha\downarrow} d_{\mathrm{i}\alpha^\prime\uparrow}
+d_{\mathrm{i}\alpha\uparrow}^\dagger d_{\mathrm{i}\alpha\downarrow}^\dagger
 d_{\mathrm{i}\alpha^\prime\downarrow} d_{\mathrm{i}\alpha^\prime\uparrow})$, it is
rather difficult to implement the HS transformation in QMC
simulation, since it would induce a rather severe sign problem even
for CPQMC method.

In order to solve this problem, we adopt a new transformation for
$e^{-\Delta\tau H_1}$, which can sufficiently suppress the sign
problem in a wide regime of parameters~\cite{sakai2004}, and develop
the two-orbital CPQMC algorithm for the $S_4$ model. In our
simulations, $e^{-\Delta\tau H_1}$ is decoupled as,

\begin{equation}\label{eqhs}
   e^{-\Delta\tau H_1}=\frac{1}{2}\sum_{\gamma=\pm1}
   e^{\lambda\gamma(f_{\mathrm{i}\uparrow}-f_{\mathrm{i}\downarrow})}
   e^{a(N_{\mathrm{i}\uparrow}+N_{\mathrm{i}\downarrow}) + bN_{\mathrm{i}\uparrow}N_{\mathrm{i}\downarrow}}
\end{equation}

with

\begin{eqnarray}\label{eqdecom}
   f_{\mathrm{i},\sigma} &=& d_{\mathrm{i},x,\sigma}^\dagger d_{\mathrm{i},y,\sigma}+
   d_{\mathrm{i},y,\sigma}^\dagger d_{\mathrm{i},x,\sigma},\\
   N_{\mathrm{\mathrm{i}},\sigma} &=& n_{\mathrm{i},x,\sigma}+n_{\mathrm{i},y,\sigma}-
   2n_{\mathrm{i},x,\sigma}n_{\mathrm{i},y,\sigma},
\end{eqnarray}
where $a$, $b$ and $\lambda$ are functions of $J$ and $\Delta\tau$,
and $\gamma=\pm1$ is the newly introduced auxiliary
field~\cite{sakai2004}. For more CPQMC calculation details for the
two-orbital model, see ref.~\cite{liu2014}.

\section{Results}
We first investigate the magnetic properties of the model at half
filling. In fig.~\ref{fig1}, the magnetic structure factor,
$S(\mathrm{k})=\frac{1}{N}\sum_{ij}
e^{i\mathrm{k}\cdot(\mathrm{r}_i-\mathrm{r}_j)}\langle(n_{\mathrm{i}\uparrow}-
n_{\mathrm{i}\downarrow})(n_{\mathrm{j}\uparrow}-n_{\mathrm{j}\downarrow})\rangle$
with $n_{i\sigma}$ being the number operator, is illustrated for
various Coulomb repulsion $U$ and the Hund's coupling $J$. From
fig.\ref{fig1}(a), we can see that $S(\pi,0)$ takes a maximum over
all the high-symmetry $k$-points along the
$(0,0)$--$(\pi,0)$--$(\pi,\pi)$--$(0,0)$, and such a maximum is
significantly enhanced on increasing $U$ with a fixed $J=0.25U$.
Similarly, with a given $U$, as shown in fig.\ref{fig1}(b), the
Hund's coupling $J$ also slightly strengthens this $(\pi,0)$ or
$(0,\pi)$ magnetic order. The property is consistent with previous
Lanczos and QMC studies~\cite{daghofer2008, moreo2009b,liu2014} for
another two-orbital model~\cite{raghu2008}.

Next we calculate the magnetic structure factor at various electron
dopings. In fig.~\ref{fig2}, three typical doping cases are plotted
for $6\times6$ and $8\times8$ lattices: the undoped $\rho_0$, the
doping density $\rho_1$ at which the system reaches the strongest
magnetic order, and the doping density $\rho_2$ near $30\%$.
Interestingly, when the system is doped away from half filling, in
both the $6\times6$ and $8\times8$ lattices we find that the
$(\pi,0)$ or $(0,\pi)$ magnetic order is manifestly favoured in the
low doping regime $(\rho_0,\rho_1)$, and then significantly
suppressed when more electrons are doped into the system. These
results also qualitatively agree with the previous QMC
study~\cite{liu2014} of the two-orbital model~\cite{raghu2008}.

\begin{figure}
\onefigure[scale=0.6]{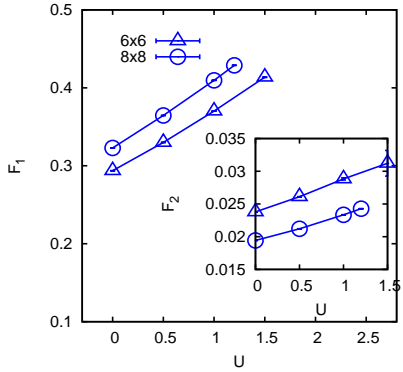}
\caption{Four-spin-operators
$F_1=\langle\vec{S}_{\mathrm{i}}^2\rangle^2-\langle(\vec{S}_{\mathrm{i}} \cdot
\vec{S}_{\mathrm{i}+\hat{x}})^2\rangle$ and
$F_2=\langle(\vec{S}_{\mathrm{i}} \cdot\vec{S}_{\mathrm{i}+\hat{x}+\hat{y}})^2\rangle-
\langle(\vec{S}_{\mathrm{i}} \cdot\vec{S}_{\mathrm{i}+\hat{x}})^2\rangle$ versus
Coulomb repulsion $U$ on $6\times 6$ and $8\times 8$ lattices in half filling with
$J=0.25U$.}
\label{fig3}
\end{figure}

\begin{figure}
\onefigure[scale=0.6]{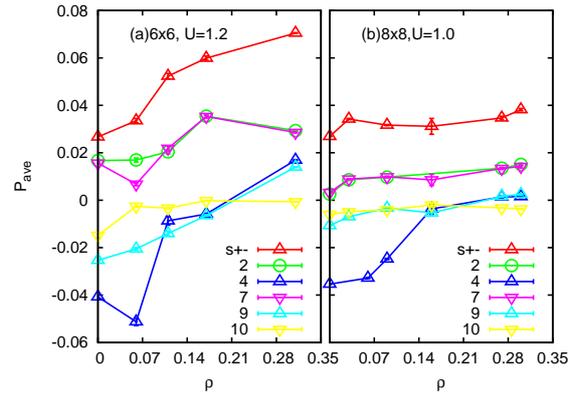} \caption{Average of long-range
pairing correlation $P_{\rm ave}$ versus various doping $\rho$ on
(a) a $6\times6$ lattice with $U=1.2$ and $J=0.25U$ and (b) an
$8\times8$ lattice with $U=1.0$ and $J=0.25U$. We follow the
classification of pairings in ref.~\cite{wan2009} with the same
meaning for the numbering.} \label{fig4}
\end{figure}

Considering the rich magnetic orders at half filling for IBSCs, we
examine the competition between orthomagnetic
(OM)~\cite{lorenzana2008} and collinear antiferromagnetic
(AFM)~\cite{dai2012} orders at half filling in the $S_4$ model.
However, the OM order, which has the nearest-neighbour magnetic
moments mutually-perpendicular with each other, behaves so similarly
with the collinear AFM order in the numerical way~\cite{moreo2009b}:
They have similar magnetic structures, almost the same expected
values of the nearest-neighbour and next-nearest-neighbour spin-spin
correlations. In order to distinguish these two magnetic orders, two
four-spin-quantities,
$F_1=\langle\vec{S}_{\mathrm{i}}^2\rangle^2-\langle(\vec{S}_{\mathrm{i}}
\cdot \vec{S}_{\mathrm{i}+\hat{x}})^2\rangle$ and
$F_2=\langle(\vec{S}_{\mathrm{i}}
\cdot\vec{S}_{\mathrm{i}+\hat{x}+\hat{y}})^2\rangle-
\langle(\vec{S}_{\mathrm{i}}
\cdot\vec{S}_{\mathrm{i}+\hat{x}})^2\rangle$, are introduced and
computed. It is argued that if the system prefers the OM phase when
increasing the Coulomb repulsion $U$, both $F_1$ and $F_2$ would go
up monotonously with $U$~\cite{liu2014}.

In fig.~\ref{fig3}, $F_1$ and $F_2$ are shown for various Coulomb
repulsion $U$ on different lattices. It is obvious that on both the
$6\times6$ and $8\times8$ lattices, $F_1$ and $F_2$ are elevated
significantly when $U$ increases, which indicts that the system
tends to be in the OM phase rather than the collinear AFM order when
the electron correlation becomes stronger. Similar results are
observed in previous QMC~\cite{liu2014} and density matrix
renormalization group~\cite{berg2010} studies.

Lastly, we discuss the pairing properties of the system. Given that
the pairing correlations within the first few distances dominate
over the long-range ones and only reflect local correlations among
spin and charge~\cite{huang2001a,huang2001b}, partial average of the
pairing correlations with distances longer than 2 lattice spacing,
$P_{\rm ave}=\frac{1}{M}\sum_{r>2}P(r)$ with $M$ being the number of
pairs and
$P(r=\left|\mathrm{i}-\mathrm{j}\right|)=\langle\Delta^{\dagger}
(\mathrm{i})\Delta(\mathrm{j})\rangle$, would be an appropriate
quantity to capture the long-range pairing properties of the system.
We mainly use $P_{\rm ave}$ to describe the pairing tendency of the
system, and for the detailed definition of
$\Delta^{\dagger}(\mathrm{i})$ for the two-orbital model, see the
discussions in refs.~\cite{liu2014,moreo2009a,moreo2009b,wan2009}.

All the possible nearest-neighbour singlet
pairings~\cite{wan2009,liu2014} and an $s_{\pm}$ channel with
next-nearest-neighbour pairing~\cite{liu2014,moreo2009b} are
calculated on $6\times6$ and $8\times8$ lattices at various dopings
and Coulomb repulsions. In fig.~\ref{fig4}, we can see that the
$s_{\pm}$-wave pairing dominates all the pairings for both the
$6\times6$ and $8\times8$ lattices under various dopings.

In addition, we find that almost all the pairings are enhanced as
more electrons are doped into the system, especially for the
$s_{\pm}$ channel. This result is different from our previous Monte
Carlo study of another two-orbital model in which the electron
doping slightly suppresses all the pairing channels~\cite{liu2014}.

Combined with the pictures of the magnetic (fig.~\ref{fig2}) and
pairing (fig.~\ref{fig4}) properties, we can hardly find an obvious
connection between the magnetic order and pairing behaviours, since
the pairing correlations are simply enhanced in the whole doping
regime, no matter whether the magnetic order is strengthened or
weakened after doping.

\section{Conclusion}
In summary, we have systemically studied the two-orbital $S_4$
symmetric microscopic model using the CPQMC method. Our simulations
demonstrate a stable $(\pi,0)$ or $(0,\pi)$ magnetic order at half
filling. Such a magnetic order is stably enhanced on increasing the
Coulomb repulsion $U$ and Hund's coupling strength $J$, which is
consistent with previous works on other two-orbital models.

Interestingly, when the system is doped away from half filling, the
magnetic order is obviously enhanced at low doping densities and
then sharply suppressed as more electrons are introduced. We also
find that the system tends to be in the OM order upon increasing
Coulomb repulsion $U$. As for the pairing properties, our
simulations strongly suggest that the $s_{\pm}$-wave pairing is the
most probable candidate.

\acknowledgments
We thank Beijing Computational Research Center for sharing the computing resources.
ZBH was supported by NSFC under Grants Nos. 11174072 and 91221103, and by SRFDP under
Grant No. 20104208110001.

\end{document}